\documentclass{elsarticle}

\usepackage{hyperref}
\usepackage{amsmath}
\usepackage{amsfonts}
\usepackage{amssymb}
\usepackage{graphicx}
\usepackage[T1]{fontenc}
\usepackage{textcomp}
\usepackage{lmodern}

\usepackage{mathdots}
\usepackage{caption}

\newtheorem{Remark}{Remark}

\newtheorem{Definition}{Definition}

\numberwithin{equation}{section}

\journal{arXiv}

\begin{document}

\begin{frontmatter}

\title{\textbf{Refined Transformation Approach for Stabilization of MIMO System by Pole~Placement}}

\author{Justin Jacob}\ead{justinjacob@iitb.ac.in}               
\author{Sreya Das}\ead{sreya\_das@iitb.ac.in}    
\author{Navin Khaneja}\ead{navinkhaneja@gmail.com}  

\address{Systems and Control Engineering Department, \\Indian Institute of Technology, Bombay}%

\begin{abstract}

The paper presents a distinctive and straightforward technique for stabilization of multi-variable systems. The idea is to decouple the system state matrix depending on different inputs and outputs. Refined special canonical transformations are described for the design of controller and observer for a single-input and single-output (SISO) case and are extended to multi-input multi-output (MIMO) systems. These transformations help in the stabilization of the error dynamics of the observer and in placing the closed loop poles of the system. The idea is not only in the transformations taken but also how the gain matrices are selected which simplifies the computation.

\end{abstract}

\begin{keyword}
controllability, cyclic subspace, Hurwitz, observability, similarity transformation, stability, state feedback.
\end{keyword}

\end{frontmatter}

\section{Introduction}

Any physical system can be expressed in the form of dynamical equations, and these govern the characteristics of the system. Around the operating point any non-linear system can be linearized. Stability and performance of a linear time-invariant (LTI) system depend on the location of closed-loop poles\cite{1698024} of the system. The stabilization of a system is done with the help of controllers where state feedback is used for placing the poles in the desired location\cite{KALMAN1960491}. The pole placement for MIMO systems is not as straight forward as the SISO case and doesn't give a unique solution. Some of the approaches mentioned in the literature\cite{1099056} include: conversion of the system to Brunovsky canonical form \cite{hermida1996brunovsky}, using Lyapnov equation to obtain the state feedback matrix without revealing the structure of the resulting feedback system subjected to the condition that state and feedback coefficient matrix has no common eigenvalues\cite{chen1999linear}, eigen structure method\cite{dooren} for solving the Sylvester's equation\cite{bhattac}, minimum number of states to obtain arbitrary pole placement by using dynamic compensators\cite{1099352}, pole placement after decoupling using Luenberger canonical form\cite{5250892}\cite{325034}. Recent studies mainly concentrate on optimizing the present techniques; Pandey\cite{7039881} presents an extensive comparison of different existing algorithms. This paper approaches the stabilization problem in a new perspective rather than going with the optimization of conventional techniques. Our idea is to decompose the state matrix to block triangular matrix form\cite{1098739} and to apply state feedback. Here a structured way for obtaining the similarity transformation is presented, which the classical literature fail to establish. In most practical cases, the states are not readily available for feedback, and we need to use a state observer. In both the controller and observer design, we try to stabilize the model by placing the poles in desired locations.

Assuming the LTI system is controllable and observable we define two special canonical transformations one for the controller and other for observer design. These similarity transformations of state matrix are refined from the controllability and observability matrices. We derive the new augmented state matrix in a lower triangular block matrix and upper triangular block matrix form for controller and observer design respectively. Initially, the approach is applied to the SISO system to get a generalized view of the components later these generalized equations are used to the diagonal blocks of the augmented system matrix of MIMO case. The transformations also give the special form of input and output matrices which helps to simplify the computations (Definition \ref{objective}). Also for reducing complexity, the gain matrix coefficients are selected in a particular way.

\begin{Definition} \label{objective}
We define the special forms of input and output matrices that will allow us to simplify the complexity in calculating the controller and observer gain matrices. In this special forms each non-zero entity corresponds to the input and output for the corresponding block of the augmented system matrix of controller and observer, respectively.\\
Special input matrix from
\begin{equation*}
	\widehat{B} = \begin{bmatrix}
   	0 & 0 & \cdots & 0\\
   	\vdots &\vdots &\cdots & \vdots\\
   	0 & 0 &\cdots & 1\\
   	\vdots &\vdots &\iddots & \vdots\\
	0 & 1 & \cdots & 0\\
	\vdots &\vdots &\cdots & \vdots\\
	1 & 0 &\cdots & 0
  	\end{bmatrix}
\end{equation*}
Special output matrix from
\begin{equation*}
	\widetilde{C} = \begin{bmatrix}
   	1 & 0 & \cdots & 0 &\cdots & \cdots & 0 & 0 & \cdots & 0\\
	0 & \cdots & 0 & 1 & 0 & \cdots & \cdots & 0 & \cdots & 0\\
	\vdots &\vdots & \vdots & \vdots & \ddots & \ddots & \vdots & \vdots & \vdots 	& \vdots \\
	0 & \cdots & \cdots & 0 & \cdots & \cdots & 1 & 0 & \cdots & 0
  	\end{bmatrix}
\end{equation*} 
This form restricts the augmented system matrix in the triangular block structure, where each diagonal block helps to attain the form similar to the SISO system.

\end{Definition}

\subsection{System Model}

Consider the $n^{th}$ order linear dynamic system given by the state and output equation

\begin{equation}\label{stateeq}
	\dot{x}=A\,x+B\,u
\end{equation}
\vspace{-.5cm}
\begin{equation}\label{outputeq}
	y=Cx
\end{equation}
where $x$ is $(n \times 1)$ state vector which belongs to the vector space $V$, $u$ is $(p \times 1)$ input vector, $y$ is $(q \times 1)$ output vector, $A$ is  $(n \times n)$ state matrix, $B$ is $(n \times p)$ input matrix and $C$ is $(q \times n)$ output matrix. An observer is used to identify the states of the system from the information of the output of the system. Model of the observer is given as

\begin{equation} \label{modelstate}
	\dot{z}=A\,z + B\,u +L(y-y_m)
\end{equation}
\vspace{-.5cm}
\begin{equation} \label{modeloutput}
	y_m = C\,z
\end{equation}
where $z$ is $(n \times 1)$ observed state vector, $y_m$ is $(q \times 1)$ observer output vector and $L$ is $(n \times q)$ observer gain matrix. The controller uses these states to stabilize the system using the state feedback. 

\begin{figure}[!ht]
\begin{center}
\includegraphics[scale=.8]{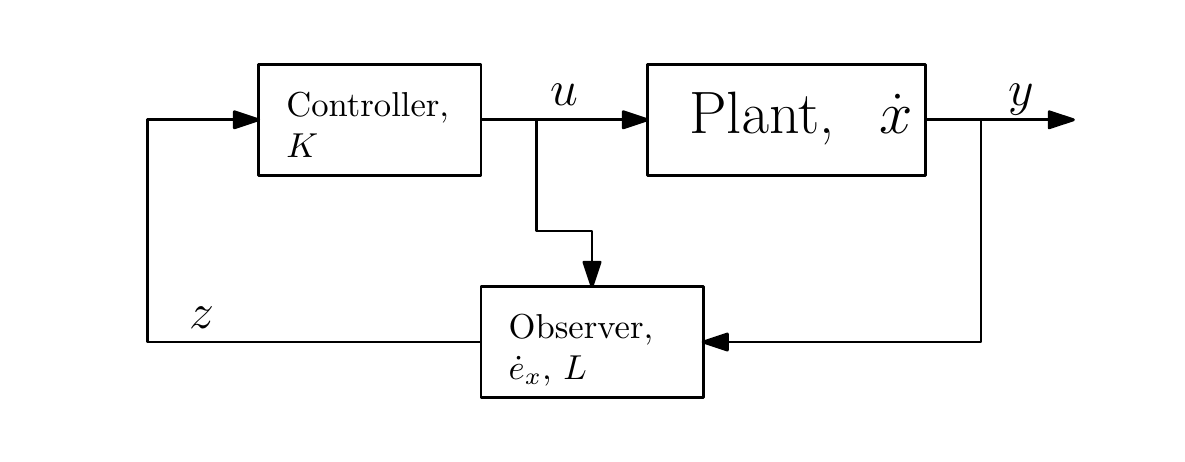}
\caption{System with observer and controller}
\label{systemfigure}
\end{center}
\end{figure}

\begin{equation} \label{controllergain}
	u = -K\,z
\end{equation}
where $K$ is $(p \times n)$ controller feedback gain matrix. Without loss of generality, here the reference signal which is kept at zero. Assuming the system is fully state controllable with the controllability matrix $M$, the column vectors will span the whole space\cite{KALMAN1960491}, $V$.

\begin{equation}\label{controllability}
	M=\begin{bmatrix}
	B & AB & A^2B & \cdots & A^{n-2}B & A^{n-1}B
	\end{bmatrix}
\end{equation}
where input matrix $B= \left[ b_1\ b_2 \ldots b_j \ldots b_p \right]$. Obtaining the cyclic subspaces $S_j$ corresponding to each input vector $b_j$'s we have $S_1 \cup S_2 \cup \cdots \cup S_j \cup \cdots \cup S_p = V$. The $dim(S_j)$ gives the controllability index of the corresponding input vector $b_j$. Assuming fully state observable, the observability matrix $N$, the row vectors will span the whole space\cite{KALMAN1960491}, $V$.

\begin{equation}
	N=\begin{bmatrix}
	C \\ C\,A \\ \vdots \\ C\,A^{n-2} \\ C\,A^{n-1}
	\end{bmatrix}
\end{equation}
where the output matrix $C=\left[c^T_1\,c^T_2...\,c^T_j\,...c^T_q\right]^T$. Obtaining the cyclic subspaces $R_j$ corresponding to each output vector $c_j$'s we have $R_1 \cup R_2 \cup \cdots \cup R_j \cup \cdots \cup R_q = V$. $dim(R_j)$ gives the observability index of the corresponding output vector $c_j$.

\section{System Stabilization}

From \eqref{stateeq} and \eqref{modelstate}, the error dynamics $\dot{e}_x$ is obtained as

\begin{align}\label{errordynamics}
	\begin{split}
	\dot{e}_x = \dot{x}-\dot{z} =& (A - LC)\,e_x \\
	y - y_m =& C\,e_x
	\end{split}
\end{align} 
Applying \eqref{controllergain} in \eqref{stateeq} along with \eqref{errordynamics} we get the total state equation as

\begin{equation}\label{totalstate}
	\begin{bmatrix}
	\dot{x} \\
	\dot{e_x}
	\end{bmatrix}
	=
	\begin{bmatrix}
	A-BK & BK \\
	0 & A-LC
	\end{bmatrix}
	\,
	\begin{bmatrix}
	x \\
	e_x
	\end{bmatrix}
\end{equation}

\begin{Remark}\textbf{
Any system of the form \eqref{totalstate} is stable iff $A-BK$ and $A-LC$ are individually Hurwitz.}\\
\end{Remark}

Since \eqref{totalstate} is of the block upper triangular form the characteristics equation is only contributed by the diagonal blocks. The eigenvalues of the observer based controller system hence are the eigenvalues of $A-BK$ and eigenvalues of $A-LC$. Thus $K$ and $L$ are designed in a way such that $A-BK$ and $A-LC$ have stable eigenvalues. By this way, we can stabilize a linear system which is controllable and observable. 

\section{Observer Design}

\subsection{Single Output Case}\label{siso_o}

Assuming the system is fully observable, $Q$ is the linear transformation to obtain the special canonical form of the system \eqref{stateeq}, where $T$ denotes the transpose. \\

\begin{Definition}\label{Form1}

\begin{equation}
	Q = \begin{bmatrix}
   	c^T & (c\,A)^T & \cdots & \cdots& (c\,{A^{n-1}})^T
 	\end{bmatrix}^T
\end{equation}
The rows are arranged from top to bottom in the ascending powers of $A$. 

\end{Definition}

The inverse exists as all the $n$ rows are linearly independent, and the transformation used is $\widetilde{z}=Q\,z$ where $\widetilde{z}^T=\begin{bmatrix} \widetilde{z}_1 & \widetilde{z}_2 & \cdots& \widetilde{z}_n\end{bmatrix}$. The transformed system equations is

\begin{equation}
	\dot{\widetilde{z}}=Q\,A\,Q^{-1}\,\widetilde{z} + Q\,b\,u +Q\,L\,(y-y_m)
\end{equation}
\vspace{-.5cm}
\begin{equation}
	y_m=c\,Q^{-1}\widetilde{z}
\end{equation}
where
\begin{equation}\label{observersingle}
	\begin{split}
	{	
	QAQ^{-1}= \begin{bmatrix}
   	0 & 1 & 0 &\cdots& 0 & 0 \\
	0 & 0 & 1 &\cdots& 0 & 0 \\
	\vdots& & &\ddots&   &\vdots& \\
	\\
	0 & 0 &\cdots&\cdots& 0 & 1 \\
	-a_1& -a_2 &\cdots&\cdots& -a_{n-1} & -a_n
	\end{bmatrix} 
	} \\\\
	\hspace{0cm}
	{
	cQ^{-1}= \begin{bmatrix}
   	1 &	0 & \cdots & & 0 & 0
	\end{bmatrix}
	}
	\end{split}
\end{equation}
$a_{n}$, $a_{n-1}$, $\cdots$, $a_1$ are the coefficient of the systems characteristic equation given by $s^{n}+a_{n}s^{n-1}+\cdots+a_{2}s+a_{1}$. The error dynamics corresponding to \eqref{errordynamics} becomes $\dot{\widetilde{e}}_x=[Q\,A\,Q^{-1}- Q\,L\,c\,Q^{-1}]\,\widetilde{e}_x$ where $\widetilde{{e}}_x = Q\,x - \widetilde{z}$. The transformed augmented state transition matrix, $G=[Q\,A\,Q^{-1}- Q\,L\,c\,Q^{-1}]$ is stabilized by $QL=\begin{bmatrix}
l_n & l_{n-1} &\cdots& l_1 \end{bmatrix}^T$. 

\[
	[sI{{-}}G]= \begin{bmatrix}
   	s+{{l}}_n & {{-}}1 & 0 &\cdots& 0 & 0 \\
	l_{n-1} & s & {{-}}1 &\cdots& 0 & 0 \\
	\vdots& &\ddots &\ddots&   &\vdots& \\
	\\
	l_{2} & 0 &\cdots&\cdots& s & {{-}}1 \\
	l_1{{+}}a_1& a_2 &\cdots&\cdots& a_{n-1} & s{{+}}a_n
	\end{bmatrix}
\]  
To get the characteristic polynomial, the determinant of above the matrix is taken with respect to the first column. 

\begin{equation}\label{polynomial}
	\begin{split}
	(s& + l_n)\left[s^{n-1}+a_n\,s^{n-2}+\cdots\cdots+a_3\,s+a_2\right]\\
	& + l_{n-1}\left[s^{n-2}+a_n\,s^{n-3}+\cdots\cdots+a_4\,s+a_3\right]\\
	&\  \vdots\\
	& + l_{2}\left[s+a_{n}\right]\\
	& + l_1+a_1
	\end{split}
\end{equation}
whose general form is 

\begin{equation}\label{generalpolynomial}
	\rho(s)+ \sum_{j=2}^{n}l_j\left(\sum_{i=1}^{j}s^{j-i}\,a_{n+2-i}\right)+ l_1
\end{equation}
where $a_{n+1}=1$ and $\rho(s)$ is the characteristic polynomial of the original system.  The coefficient of $s^{n-1}$ is a function of $l_n$ and $a_n$, where $a_n$ is the system coefficient and from the knowledge of the required eigenvalue to stabilize the error dynamics, we can obtain the value of $l_n$. All the constant terms are compared to the desired characteristic equation $s^{n}+\alpha_n\,s^{n-1}+\cdots\cdots+\alpha_2\,s+\alpha_1$

\begin{equation}
	\begin{aligned}	
	s^{n-1}\Rightarrow & \alpha_n\Longrightarrow l_n+a_n \\
	s^{n-2}\Rightarrow & \alpha_{n-1}\Longrightarrow l_{n-1}+l_n\,a_n+a_{n-1} \\
	\vdots \\
	s^1\Rightarrow & \alpha_2\Longrightarrow l_{2}+l_{3}\,a_n+\cdots + l_n\,a_3+a_2 \\
	s^0\Rightarrow & \alpha_1\Longrightarrow l_1+l_{2}\,a_n+\cdots +l_n\,a_2+a_1 \\
	\end{aligned}
\end{equation}
The generalized coefficient of the augmented characteristic polynomial is 

\begin{equation}\label{generalcoefficentobserver}
	\alpha_i=\sum_{j=i}^{n+1}l_j\, a_{n+1+i-j}
\end{equation}
where $a_{n+1}=l_{n+1}=1$ and $\alpha_i$ are the desired coefficient for stabilizing observer error dynamics. 

\subsection{Multiple Output Case}

System \eqref{outputeq} with $q$ outputs has the output matrix $C=\left[c^T_1\,c^T_2...\,c^T_j\,...c^T_q\right]^T$. \\

\begin{Definition}

Refined transformation matrix $Q$ is obtained from the corresponding basis of the constrained cyclic subspaces, $U_j$ generated by $c_j$. 
\begin{equation}\label{eq:o3}
	\begin{aligned}
	Q=[\,{c_1}^T \cdots ({c_1} {A^{m_{1}-1}})^T| \cdots\, |\,{c_j}^T\cdots ({c_j} 		{A^{m_{j}-1}})^T\,| \\ \cdots|{c_k}^T \,\cdots \,({c_k} {A^{m_{k}-1}})^T \,]^T
	\end{aligned}
\end{equation}
Here the subspaces are subjected to the condition where the total space $V = U_1 \oplus U_2 \oplus \cdots \oplus U_j \oplus \cdots \oplus U_k$, $k\leq q$ where $dim(U_j) = m_j$ and $\sum_{j=1}^{k}m_{j} = dim(V) = n$. The arrangement is similar to \textit{Definition} \ref{Form1}.

\end{Definition}

$Q$ transforms the system matrix to block upper triangular matrix and the output matrix to the special canonical form in Definition \ref{objective}.

\begin{equation}\label{eq:o6}
	\begin{split}
	{
	QAQ^{-1} = \begin{bmatrix} \widetilde{A}_1 & \bigstar & \bigstar &\cdots & \bigstar\\ 
	0 & \widetilde{A}_2 & \bigstar & \cdots &  \bigstar \\
	\vdots & \vdots & \widetilde{A}_j & \cdots & \vdots \\
	\vdots & \vdots & \vdots & \ddots & \bigstar \\
	0 & 0 & 0 & \cdots &  \widetilde{A}_{k}
	\end{bmatrix}
	}
	\hspace{0cm} \\ \\
	{
	CQ^{-1} = \begin{bmatrix}
   	\widetilde{C}_1 & 0 &\cdots & 0\\
	0 & \widetilde{C}_2 & \cdots & 0\\
	\vdots &\vdots & \widetilde{C}_j & \vdots\\
	0 & 0 &\cdots & 0\\
	0 & 0 &\cdots & \widetilde{C}_k
	\end{bmatrix}
	}	
	\end{split}
\end{equation}
where $\widetilde{A}_j$ is $(m_j \times m_j)$ and $\widetilde{C}_j$ is $(1 \times m_j)$  matrices which corresponds to the form \eqref{observersingle}. $\bigstar$ denotes $\widetilde{A}_{ij}$ and has a dimension of $m_i \times m_j$, which does not involves in the calculation. \\

\begin{Definition}

$\widetilde{L}$ to stabilize the error dynamics has $m_j$ number of non zero coefficients in each column corresponding to $c_j$ in the $Q$ matrix. 

\begin{equation}\label{observergain}
	\widetilde{L} = \begin{bmatrix} 
	l^1_{m_1} & 0 & \cdots & \cdots & 0 \\
	\vdots & \vdots & & \vdots & \vdots \\
	l_{1}^1 & 0 & \cdots & \cdots & \vdots\\
	0 & l_{m_j}^j & \ddots & \cdots & \vdots \\
	\vdots & \vdots &  & \vdots & \vdots \\
	0 & l_{1}^j & \ddots & \cdots & \vdots\\
	\vdots & \vdots & \ddots & \vdots & \vdots \\
	0 & 0 & \ddots & \cdots & l_{m_k}^k \\
	\vdots & \vdots & & \vdots & \vdots \\
	0 & 0 & \cdots & \cdots & l_1^k
	\end{bmatrix}	
\end{equation}

\end{Definition}

If the $Q$ matrix has used only $k$ output vector from $C$ matrix, where $k<q$ then, $q-k$ columns of the $\widetilde{L}$ matrix all takes zeros as it has dimension $(n \times q)$. The choice of $\widetilde{L}$ will preserve the block upper triangular form in the augmented system matrix $\widetilde{A} - \widetilde{L} \widetilde{C}$, where  $\widetilde{A}=QAQ^{-1}$, $\widetilde{C}=CQ^{-1}$ and $\widetilde{L}= QL$.\\

The characteristic matrix $\left[ sI-\left( \widetilde{A} - \widetilde{L} \widetilde{C} \right) \right] $ has each diagonal block similar to the single output case. $det(sI-\widetilde{A} + \widetilde{L} \widetilde{C})$ is obtained from the product of diagonal blocks. \textit{Note for the untransformed system the observer gain $L = Q^{-1} \widetilde{L}$}.
 
\begin{equation}\label{decoupleeq}
	\left|(sI_{m_1}-\widetilde{A}_1)\right|* \cdots *\left|(sI_{m_j}-\widetilde{A}_{j})		\right|*\cdots *\left|(sI_{m_k}-\widetilde{A}_k)\right|
\end{equation} 
Each block has an order $m_j$ and can be separately evaluated as a single output case discussed in the previous section \ref{siso_o}. The desired values of the poles say ($\gamma_1, \gamma_2, \cdots, \gamma_n$), to eliminate the error, as quickly as possible, is placed with the help of each block. $j^{th}$ block helps in placing $m_j$ poles.

\section{Controller Design}

\subsection{Single Input Case}

Assuming the system is fully controllable, we take $P$ as the linear transformation to obtain the special canonical form for the system \eqref{stateeq}. \\

\begin{Definition}\label{Form2}

\begin{equation}\label{Controllability}
	P=\begin{bmatrix}
	A^{n-1}\,B & A^{n-2}\,B & \cdots & \cdots& A\,B & B
	\end{bmatrix}
\end{equation} 
The columns are arranged from left to right in the ascending powers of $A$. 

\end{Definition}

Since $P$ has $n$ linearly independent columns, it has a rank $n$, and its range is the whole space\cite{KALMAN1960491} $V$. The transformed state, $\widehat{x}=P^{-1}\,x$, where $\widehat{x}^T=\begin{bmatrix} \widehat{x}_1 & \widehat{x}_2 & \cdots& \widehat{x}_n\end{bmatrix}$, and the transformed system equation is 

\begin{equation}
	\dot{\widehat{x}}=P^{-1}\,A\,P\,\widehat{x}+P^{-1}\,B\,u
\end{equation}
 where,
\begin{equation}\label{controllersingle}
	\begin{split}
	{
	P^{-1}\,A\,P= \begin{bmatrix}
   	-a_n & 1 & 0 &\cdots& 0 & 0 \\
	-a_{n-1} & 0 & 1 &\cdots& 0 & 0 \\
	\vdots& & &\ddots&   &\vdots& \\
	\\
	-a_2 & 0 &\cdots&\cdots& 0 & 1 \\
	-a_1& 0 &\cdots&\cdots& 0 & 0
	\end{bmatrix}
	}
	\hspace{0cm} \\ \\
	{
	P^{-1}\,B= \begin{bmatrix}
   	0\\
	0\\
	\vdots\\
	\\
	0\\
	1
	\end{bmatrix}
	}
	\end{split}
\end{equation}
$a_{n}$, $a_{n-1}$, $\cdots$, $a_1$ are the coefficient of the systems characteristic equation. Using the estimated state vector $z$, the \textbf{feedback law} used is $u=-K\,z$, where $K=\begin{bmatrix}
k_1 &  k_2 &\cdots& k_n \end{bmatrix}$. After transformation two terms obtained in the system equation, are $\widehat{x}$ and $\widehat{e}_z = \widehat{x}- \widehat{z}$, where $\widehat{z}=P^{-1}z$. The total system dynamics is given by

\begin{equation}\label{controllertotalstate}
	\begin{bmatrix}
	\dot{\widehat{x}} \\
	\dot{\widehat{e}_x}
	\end{bmatrix}
	=
	\begin{bmatrix}
	\widehat{A}-\widehat{B}\widehat{K} & \widehat{B}\widehat{K} \\
	0 & P^{-1}(A-LC)P
	\end{bmatrix}
	\,
	\begin{bmatrix}
	\widehat{x} \\
	\widehat{e}_x
	\end{bmatrix}
\end{equation}
where $\widehat{A}=P^{-1}AP$, $\widehat{B}=P^{-1}B$ and $\widehat{K}= KP$.
Since the lower block of \eqref{totalstate} and \eqref{controllertotalstate} are similar, we need to consider only the first block of the total system as we have already taken care of the lower part in the observer design. So the augmented state transition matrix required is $F=[\widehat{A}-\widehat{B} \widehat{K}]$ where $\widehat{K} =\begin{bmatrix}
\widehat{k}_1 & \widehat{k}_2 &\cdots& \widehat{k}_n \end{bmatrix}$. Now the characteristic polynomial is obtained and has a form similar to \eqref{polynomial} where $a_j$ is in place of $l_j$ and $\widehat{k}_j$ in place of $a_j$. We use the same \eqref{generalpolynomial} with the corresponding changes, where $\widehat{k}_{n+1} =1 $. Let \{$\lambda_1, \lambda_2, \cdots, \lambda_n$\} is the desired set of poles of the system and say the desired characteristic polynomial to be $s^{n}+\beta_n\,s^{n-1}+\cdots\cdots+\beta_2\,s+\beta_1$. The general coefficient form is

\begin{equation}\label{generalcoefficentcontroller}
	\beta_i=\sum_{j=i}^{n+1}a_j\, \widehat{k}_{n+1+i-j}
\end{equation}
where $a_{n+1}=\widehat{k}_{n+1}=1$. Coefficients $\widehat{k}_j$ are obtained similar to the observer case and these stabilizes the closed loop system.

\subsection{Multiple Input Case}
System \eqref{stateeq} with $p$ inputs has the input matrix $B= \left[ b_1\ b_2 \ldots b_j \ldots b_p \right]$. \\

\begin{Definition}

Refined transformation matrix $P$ is obtained from the corresponding basis of the constrained cyclic subspaces ($W_j$) generated by $b_j$.
\begin{equation}\label{eq:m3}
	\begin{aligned}
	P=[ A^{n_{k}-1}b_k\,\cdots b_k | \cdots | \,A^{n_{j}-1}b_j \cdots b_j\,| \\			\cdots|\, A^{n_{1}-1}b_1 \cdots b_1]
	\end{aligned}
\end{equation}
Here the subspaces are subjected to the condition where the total space $V = W_1 \oplus W_2 \oplus \cdots \oplus W_j \oplus \cdots \oplus W_k$, $k\leq p$ where $dim(W_j) = n_j$ and $\sum_{j=1}^{k}n_{j} = dim(V) = n$. The arrangement is similar to \textit{Definition} \ref{Form2}.

\end{Definition}

$P$ transforms the system matrix to block lower triangular matrix and the input matrix to the special canonical form in Definition \ref{objective}.

\begin{equation}\label{eq:m6}
	\begin{split}
	{
	P^{-1}AP = \begin{bmatrix} \widehat{A}_{k} & 0 & 0 &\cdots & 0\\ 
	\star & \widehat{A}_{k-1} & 0 & \cdots &  0 \\
	\vdots & \ddots & \widehat{A}_j & \cdots & \vdots \\
	\vdots & \vdots & \ddots & \ddots & \vdots \\
	\star & \star & \star & \cdots &  \widehat{A}_{1}
	\end{bmatrix}
	}
	\hspace{0cm} \\ \\
	{
	P^{-1}\,B= \begin{bmatrix}
   	0 & 0 &\cdots & \widehat{B}_k\\
	0 & 0 & \cdots & 0\\
	\vdots &\vdots & \widehat{B}_j & \vdots\\
	0 & \widehat{B}_2 &\cdots & 0\\
	\widehat{B}_1 & 0 &\cdots & 0
	\end{bmatrix}
	}
	\end{split}
\end{equation}
where $\widehat{A}_j$ is $(n_j \times n_j)$ and $\widehat{B}_j$ is $(n_j \times 1)$  matrices which correspond to the form \eqref{controllersingle}. $\star$ denotes $\widehat{A}_{ij}$ and has a dimension of $n_i \times n_j$, which does not involve in the calculation. \\

\begin{Definition}

$\widehat{K}$ which places the closed loop poles has $n_j$ number of non zero coefficients in each rows corresponding to $b_j$ in the $P$ matrix. 

\begin{equation}\label{eq:m8}
	\widehat{K} = \begin{bmatrix} 
	0 &\cdots & 0 & 0 &\cdots & 0 &\widehat{k}_{1}^1 &\cdots &\widehat{k}_{n_{1}}^1 \\
	0 &\cdots & 0 &\widehat{k}_{1}^j &\cdots &\widehat{k}_{n_{j}}^j & 0 &\cdots & 0 \\
	\vdots &\cdots &\vdots &\vdots &\cdots &\vdots &\vdots &\cdots &\vdots \\
	\widehat{k}_{1}^k &\cdots& \widehat{k}_{n_{k}}^k &0 &\cdots & 0 &\cdots &\cdots & 0
	\end{bmatrix}	
\end{equation}
\end{Definition}

The choice of $\widehat{K}$ will preserve the block lower triangular form in the augmented system matrix $[\widehat{A} - \widehat{B} \widehat{K}]$. The characteristic matrix $\left[ sI-\left( \widehat{A} - \widehat{B} \widehat{K} \right) \right] $ has each diagonal block similar to the single input case. The $det(sI-\widehat{A} + \widehat{B} \widehat{K})$ is obtained from the product of diagonal blocks. \textit{Note for the untransformed system the feedback gain $K =  \widehat{K} P^{-1}$ }. The desired values of the closed loop poles to stabilize the system can be placed by each block individually, similar to single input case. $L$ and $K$ stabilizes the system \eqref{totalstate}. \\

\textit{In order to have a faster response, the poles of the observer can be taken more negative w.r.t the controller}. It may be noted that the column vectors of input and row vectors of output matrix can be taken in any sequential order and there will be a corresponding change in the transformation. Nevertheless, the forms obtained for the system matrices will follow the defined ones. 

\section{Conclusions}

This paper presents a simple and straight forward approach towards the stabilization of linear time-invariant MIMO systems. Although the controllability and observability transformation approach are widely used, the refined transformation approach simplifies the problem prominently. The defined similarity transformations help to obtain the special canonical forms of the input and output matrix (Definition \ref{objective}). This transformation transforms the state matrix to the block triangular form, which enables to obtain similar characteristic polynomial for controller and observer case. Besides, the controller and observer gain matrices are defined in such a way that reduces the computational complexity in calculating the gain matrix coefficients. The characteristic polynomial for MIMO case corresponds to the product of the diagonal blocks in the augmented state matrix which resembles the SISO case. The significant achievement is that it follows a similar equation which satisfies for both observer and controller design. It must be noted that if input order is changed, a new input matrix corresponding to the change is obtained. This changes the special transformation matrix and enables us to use different inputs to control the system so that the designer can optimize the use of control effort corresponding to different inputs. The same can be done in the case for observer design with rearranging the output.
    
\bibliography{reference} 

\begin{thebibliography}{10}
\expandafter\ifx\csname url\endcsname\relax
  \def\url#1{\texttt{#1}}\fi
\expandafter\ifx\csname urlprefix\endcsname\relax\def\urlprefix{URL }\fi
\expandafter\ifx\csname href\endcsname\relax
  \def\href#1#2{#2} \def\path#1{#1}\fi

\bibitem{1698024}
J.~H. {Mulligan}, The effect of pole and zero locations on the transient
  response of linear dynamic systems, Proceedings of the IRE 37~(5) (1949)
  516--529.
\newblock \href {https://doi.org/10.1109/JRPROC.1949.232649}
  {\path{doi:10.1109/JRPROC.1949.232649}}.

\bibitem{KALMAN1960491}
R.~Kalman, On the general theory of control systems, IFAC Proceedings Volumes
  1~(1) (1960) 491 -- 502, 1st International IFAC Congress on Automatic and
  Remote Control, Moscow, USSR, 1960.
\newblock \href {https://doi.org/https://doi.org/10.1016/S1474-6670(17)70094-8}
  {\path{doi:https://doi.org/10.1016/S1474-6670(17)70094-8}}.

\bibitem{1099056}
E.~Davison, W.~Wonham, On pole assignment in multivariable linear systems, IEEE
  Transactions on Automatic Control 13~(6) (1968) 747--748.
\newblock \href {https://doi.org/10.1109/TAC.1968.1099056}
  {\path{doi:10.1109/TAC.1968.1099056}}.

\bibitem{hermida1996brunovsky}
J.~A. Hermida-Alonso, M.~P. Perez, T.~Sanchez-Giralda, Brunovsky's canonical
  form for linear dynamical systems over commutative rings, Linear Algebra and
  its applications 233 (1996) 131--147.

\bibitem{chen1999linear}
C.-T. Chen, Linear system theory and design (1999).

\bibitem{dooren}
P.~{Van Dooren}, The generalized eigenstructure problem in linear system
  theory, IEEE Transactions on Automatic Control 26~(1) (1981) 111--129.
\newblock \href {https://doi.org/10.1109/TAC.1981.1102559}
  {\path{doi:10.1109/TAC.1981.1102559}}.

\bibitem{bhattac}
S.~Bhattacharyya, E.~de~Souza, Pole assignment via sylvester's equation,
  Systems \& Control Letters 1~(4) (1982) 261 -- 263.
\newblock \href {https://doi.org/https://doi.org/10.1016/S0167-6911(82)80009-1}
  {\path{doi:https://doi.org/10.1016/S0167-6911(82)80009-1}}.

\bibitem{1099352}
F.~{Brasch}, J.~{Pearson}, Pole placement using dynamic compensators, IEEE
  Transactions on Automatic Control 15~(1) (1970) 34--43.
\newblock \href {https://doi.org/10.1109/TAC.1970.1099352}
  {\path{doi:10.1109/TAC.1970.1099352}}.

\bibitem{5250892}
P.~K. Kar, J.~D. Aplevich, N.~J. Bergman, Decouplable multivariable systems:
  canonical forms and pole assignment, Electrical Engineers, Proceedings of the
  Institution of 120~(11) (1973) 1433--1438.
\newblock \href {https://doi.org/10.1049/piee.1973.0291}
  {\path{doi:10.1049/piee.1973.0291}}.

\bibitem{325034}
M.~Valasek, N.~Olgac, Generalization of ackermann's formula for linear mimo
  time invariant and time varying systems, in: Proceedings of 32nd IEEE
  Conference on Decision and Control, 1993, pp. 827--832 vol.1.
\newblock \href {https://doi.org/10.1109/CDC.1993.325034}
  {\path{doi:10.1109/CDC.1993.325034}}.

\bibitem{7039881}
A.~{Pandey}, R.~{Schmid}, T.~{Nguyen}, Y.~{Yang}, V.~{Sima}, A.~L. {Tits},
  Performance survey of robust pole placement methods, in: 53rd IEEE Conference
  on Decision and Control, 2014, pp. 3186--3191.
\newblock \href {https://doi.org/10.1109/CDC.2014.7039881}
  {\path{doi:10.1109/CDC.2014.7039881}}.

\bibitem{1098739}
W.~{Wonham}, On pole assignment in multi-input controllable linear systems,
  IEEE Transactions on Automatic Control 12~(6) (1967) 660--665.
\newblock \href {https://doi.org/10.1109/TAC.1967.1098739}
  {\path{doi:10.1109/TAC.1967.1098739}}.

\end{thebibliography}
\bibliographystyle{elsarticle-num}

\end{document}